\documentclass[reprint,a4paper,aps,prl,amsmath,amsfonts,showpacs,floatfix,nofootinbib]{revtex4-1}
\usepackage{bm}                      
\usepackage{graphicx}                
\usepackage{dcolumn}                 
\usepackage[table]{xcolor}
\usepackage{amsmath}

\newcommand*{\cent}[1]{\multicolumn{1}{c}{$#1$}}
\newcolumntype{w}[1]{D{.}{.}{#1}}

\newcommand{\icm}{\text{cm}^{-1}}
\newcommand{\br}{\vec{r}}

\newcommand{\me}{m_{\mathrm{e}}}

\newcommand{\SE}{Schr{\"o}dinger equation}
\newcommand{\riA}{r_{1A}}
\newcommand{\rjA}{r_{2A}}
\newcommand{\riB}{r_{1B}}
\newcommand{\rjB}{r_{2B}}
\newcommand{\rij}{r_{12}}

\begin{document}


\title{\SE\ solved for the hydrogen molecule with unprecedented accuracy}

\author{Krzysztof Pachucki}
\email{krp@fuw.edu.pl}
\affiliation{Faculty of Physics,
             University of Warsaw, Pasteura 5, 02-093 Warsaw, Poland}

\author{Jacek Komasa} 
\email{komasa@man.poznan.pl}
\affiliation{Faculty of Chemistry, Adam Mickiewicz University,
             Umultowska 89b, 61-614 Pozna{\'n}, Poland}

\date{\today}

\begin{abstract}
The hydrogen molecule can be used for determination of physical constants and for improved
tests of the hypothetical long range force between hadrons, which requires a sufficiently accurate knowledge
of the molecular levels. For this reason, we have undertaken a project of significant improvements 
in theoretical predictions of H$_2$ and perform the first step, which is the solution of the nonrelativistic 
\SE\ to the unprecedented accuracy of $10^{-12}$. This will inspire, we hope, a parallel progress 
in the spectroscopy of the molecular hydrogen. 

\end{abstract}

\maketitle


\section{Introduction}

The spectroscopy of simple atomic systems like hydrogen \cite{Parthey:11,Schwob:99}, 
hydrogenic ions \cite{Sturm:14}, muonic hydrogen \cite{Pohl:10}, muonium, positronium 
has been used to determine fundamental physical constants and to test the quantum electrodynamic theory.
Although experiments for more complicated atomic systems like helium or lithium can be as accurate as 
for hydrogen, the precision of theoretical predictions, at the moment, is not sufficient to determine 
physical constants, such as the fine structure constant $\alpha$, the Rydberg constant (Ry), 
or the absolute value of the nuclear charge radius. In contrast, the hydrogen molecule, 
thanks to its simplicity, has already been used for the most accurate determination 
of the deuteron magnetic moment from NMR measurements \cite{PKP:15}, and for the studies 
of the unknown hypothetical fifth force at the atomic scale \cite{Ubachs:13}. 
The precision achieved recently for transition frequencies, of an order of $10^{-4}\,\icm$, 
has been verified by a series of measurements, which resulted in strong bounds
on the fifth force. We will argue in this work, that it is possible to achieve $10^{-6}\,\icm$ 
accuracy for energy levels of the hydrogen molecule, which not only will improve tests 
of quantum electrodynamics theory and will put stronger bounds on the fifth force, but also 
will allow a resolution of the proton charge radius puzzle, which stands as a violation 
of the Standard Model of fundamental interactions \cite{Pohl:13}.      

\begin{figure}[!hbt]
\includegraphics[scale=0.65]{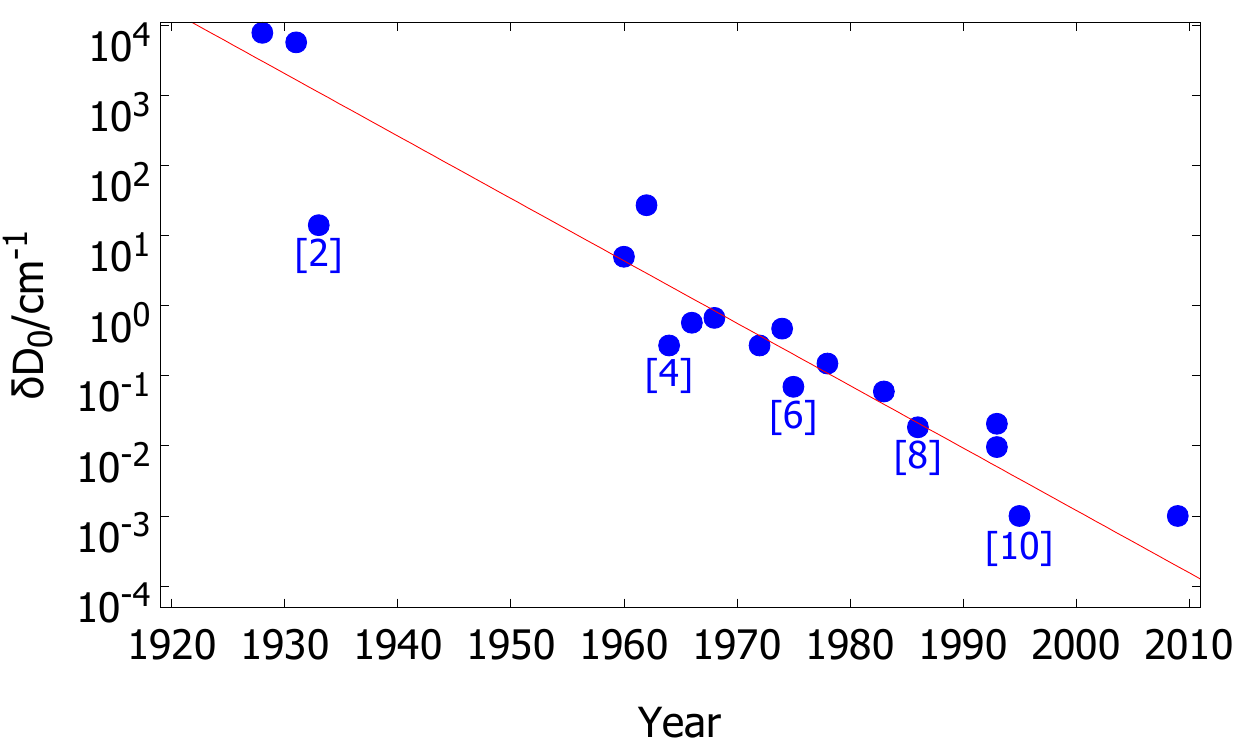}
\caption{The accuracy of theoretical predictions of dissociation energy $D_0$ of H$_2$ versus time,
with the linear fit on the logarithmic scale.}
\label{Fig}
\end{figure}

The improvement in theoretical predictions for the hydrogen molecule can be achieved 
by the calculation of the yet unknown higher order $\alpha^4$\,Ry quantum electrodynamics 
correction and by a more accurate solution of the nonrelativistic \SE. 
This second improvement is performed in this work, while the calculation of the QED effects is in progress.     
The solution of the \SE\ for the hydrogen molecule
has been pursued since almost the beginning of the quantum mechanics theory.
Over time, there have been many contributions to the development of methods 
with increased precision of theoretical predictions.
Heitler and London \cite{Heitler:27}, James and Coolidge \cite{James:33}, 
Ko{\l}os and Wolniewicz \cite{Kolos:64a}, 
and many others have made their marks on the history of research on H$_2$.
Every breakthrough in the precision of theoretical predictions has been related to 
the progress in computational techniques.
Fig.~\ref{Fig} illustrates the progress made over many decades
in the precision of the dissociation energy $D_0$ for H$_2$. 

Not always the results of calculations have been in agreement with measured values, which has questioned 
the validity of the theoretical approach. For example, in 1964 Ko{\l}os and Wolniewicz \cite{Kolos:64b} 
solved variationally the nonadiabatic \SE\ for the hydrogen molecule. The calculated dissociation energy 
appeared to be higher than the measured one \cite{Herzberg:60}, which was in contradiction with 
the variational principle. Five years later, Herzberg measured $D_0$ again 
with the accuracy increased to a few tenths of reciprocal centimeter and obtained
a value higher than previously by $5\,\icm$ in agreement with the Ko{\l}os and Wolniewicz predictions.

Further progress in theoretical predictions was related to the calculations of the leading relativistic 
corrections and to approximate QED effects \cite{Kolos:75,Wolniewicz:83,Kolos:86,Wolniewicz:93,Wolniewicz:95}.  
Later on, due to rapid development of computer power, the methods based on exponentially correlated 
Gaussian (ECG) functions have been developed both in the Born-Oppenheimer approximation \cite{PK09} 
and in the direct nonadiabatic approach \cite{Bubin:03,Bubin:09b}. 
Very recently, we have introduced a nonadiabatic perturbation theory (NAPT), which  
allowed the accuracy of about $10^{-3}\,\icm$ to be achieved for all the rovibrational 
levels of H$_2$ and isotopomers \cite{PK09}. 
However, the complexity of NAPT in the higher order of electron-nucleus mass ratio \cite{PK15}
makes further improvements in accuracy quite complicated. For this reason, we propose
another approach to the direct solution of the \SE\ by the use
of the special integration technique for explicitly correlated exponential functions and
present its first results in this Letter. 
Over fifty years after Ko{\l}os and Wolniewicz, we approach the problem of solving
the four-body \SE, with the precision aim for $D_0$ at the level of $10^{-7}\,\icm$.

\section{Theory}

The main purpose of this work is to solve accurately the stationary \SE\ $\hat{H}\Psi=E\Psi$
for a  diatomic molecule with the nuclei of charge $Z_A$ and $Z_B$
and finite masses $M_A$ and $M_B$
\begin{align}
\hat{H}=&-\frac{1}{2\,M_A}\nabla_{\!A}^2-\frac{1}{2\,M_B}\nabla_{\!B}^2
        -\frac{1}{2\,\me}\nabla_1^2-\frac{1}{2\,\me}\nabla_2^2 \nonumber\\
        &+\frac{Z_A\,Z_B}{r_{AB}}+\frac{1}{\rij}
        -\frac{Z_A}{\riA}-\frac{Z_A}{\rjA}-\frac{Z_B}{\riB}-\frac{Z_B}{\rjB}\,,
\end{align}
using the variational approach. The trial wave function
\begin{equation}\label{Expand}
\Psi(\br_1,\br_2,\vec{R}_A,\vec{R}_B)=
\sum_{k=1}^K c_k\,\hat{S}\,\psi_{\{k\}}(\br_1,\br_2,\vec{R}_A,\vec{R}_B)
\end{equation}
is expanded in properly symmetrized ($\hat{S}$), four-particle basis of exponential functions
\begin{align}
\psi_{\{k\}}
&=\exp{\left[-\alpha\,r_{AB}-\beta\,(\zeta_1+\zeta_2)\right]}
r_{AB}^{k_0}\,\rij^{k_1}\,\eta_1^{k_2}\,\eta_2^{k_3}\,\zeta_1^{k_4}\,\zeta_2^{k_5}\,,
\end{align}
where $\zeta_i=r_{iA}+r_{iB}$ and $\eta_i=r_{iA}-r_{iB}$ are coordinates closely related 
to the prolate spheroidal coordinates of $i$-th electron, $r_{ij}$ are interparticle
distances, $\alpha$ and $\beta$ are nonlinear variational
parameters, and $k_i$ are non-negative integers collectively denoted as $\{k\}$.
For its resemblance to the electronic James-Coolidge function, we call this basis function
{\em the nonadiabatic James-Coolidge} (naJC) function.

Application of the naJC function for evaluation of the matrix elements leads to 
a certain class of integrals. Efficient evaluation of these integrals has become 
feasible since the discovery of the analytic formulas \cite{Fromm:87,Harris:97}
and the corresponding recursive relations 
\cite{Pachucki:09,Pachucki:12b}. For example, the master integral, which is the starting point
for the recursions, can be expressed in the following form
\begin{align}
\int\frac{dV}{(4\,\pi)^3}&
\frac{
e^{-t\,r_{AB}}\,e^{-u\,(r_{1A}+r_{1B})}\,e^{-w\,(r_{2A}+r_{2B})}
}{r_{AB}\,\rij\,\riA\,\riB\,\rjA\,\rjB}
\nonumber \\ =&\frac{1}{4\,u\,w}\,\Biggl[
-\frac{\ln \left(\frac{2\,u}{t+u+w}\right)}{t-u+w}
-\frac{\ln \left(\frac{2\,w}{t+u+w}\right)}{t+u-w}
\nonumber \\&\quad
+\frac{\ln\left(\frac{2\,u\,w}{(u+w) (t+u+w)}\right)}{t+u+w}
+\frac{\ln \left(\frac{2\,(u+w)}{t+u+w}\right)}{t-u-w}
\Biggr]\,.
\end{align}
The integrals with additional positive powers of interparticle distances
are obtained by straightforward algebraic recursion relations,
which nevertheless are two long to be written explicitly here.
All of them can be derived from a single fourth-order differential equation \cite{Pachucki:12b},
which is satisfied by the general four-body integral,
and are expressed in terms of logarithmic and rational functions.
Details of this recursion method will be described elsewhere.

\section{Numerical results}

Results of our calculations for H$_2$ are presented in Tab.~\ref{Tconv} in the form
a sequence of energies resulting from increasing length ($K$) of expansion~(\ref{Expand}). 
The selection of $K$ was made on the basis of the saturation 
of consecutive 'shells' limited by $\sum_{k=1}^5 k_i\le\Omega$ with $k_0$ fixed at 30.
The observed regular convergence, obeying the inverse power low, permits a firm extrapolation
to the complete basis set as well as an estimation of the uncertainty.
The final value agrees well with the previous estimation of $-1.164\,025\,030\,84(6)$
a.u. obtained by Bubin et al. \cite{Bubin:09b} but has a significantly smaller uncertainty.


\begin{table}[!htb]
\caption{Convergence of the \SE\ eigenvalue $E$ (in a.u.) and of the corresponding dissociation energy $D_0$
(in $\icm$) for H$_2$ with the size of the basis set.}
\label{Tconv}
\begin{ruledtabular}
\begin{tabular*}{\textwidth}{c@{\ }@{\extracolsep{\fill}}rw{4.17}w{7.12}}
$\Omega$ & $K\quad$ & \cent{E\qquad} & \cent{D_0\qquad}  \\
\hline\\[-8pt]
      10 &  36642 & -1.164\,025\,030\,821\,4    & 36\,118.797\,732\,57 \\
      11 &  53599 & -1.164\,025\,030\,870\,9    & 36\,118.797\,743\,43 \\
      12 &  76601 & -1.164\,025\,030\,880\,4    & 36\,118.797\,745\,52 \\
      13 & 106764 & -1.164\,025\,030\,882\,5    & 36\,118.797\,745\,97 \\
$\infty$ &$\infty$& -1.164\,025\,030\,884(1)    & 36\,118.797\,746\,3(2) \\
\end{tabular*}
\end{ruledtabular}
\end{table}

%

Further increase in the accuracy of eigenvalue of the four-body \SE\ is feasible,
but the problem we face is the lack of the parallel code in multiprecision arithmetics
for the $LDL^T$ matrix decomposition with pivoting, which results in a long computation time.
However, current uncertainties in the electron-proton (proton-deuteron) mass ratio
and in the Rydberg constant are much more significant than those due to numerical uncertainties.
For example, the CODATA 2014 \cite{CODATA:14} electron-proton mass ratio has a relative uncertainty of
$9.5\cdot 10^{-11}$, which affects the eigenvalue of H$_2$ at the level of $4.3\cdot 10^{-12}$ a.u.
and the corresponding dissociation energy at $8.5\cdot 10^{-7}\,\icm$.
Similarly, the current uncertainty in the Rydberg constant affects the conversion
of $D_0$ value from a.u. to reciprocal centimeters at the level of $2.1\cdot 10^{-7}\,\icm$.
This indicates, that one cannot exclude the possibility, to determine the electron-proton mass ratio
from future high precision studies of H$_2$.

\section{Conclusions}

The approach based on explicitly correlated exponential functions
and the obtained results pave the way to a significant progress in the theory of the hydrogen 
molecule. A similar precision for nonrelativistic energies can be achieved for 
any other molecular level of H$_2$, D$_2$, HD, and HeH$^+$. 
Considering the leading relativistic corrections, they can be expressed in terms of an expectation 
value with the nonrelativistic wave function, so their evaluation does not pose a significant problem,
and they have already been calculated in the BO approximation. The leading quantum electrodynamics effects
are more complicated due to Bethe logarithm contribution, which involves
the logarithm of the nonrelativistic Hamiltonian. Its calculation beyond the BO approximation
might be problematic. However, much more challenging is the calculation of the higher order 
$\alpha^4$\,Ry contribution, which apart from hydrogen, was calculated only for He atom. 
Its magnitude can be estimated
by $2\,\alpha^2$ times the known leading relativistic correction to $D_0$ of $0.5\;\icm$, 
which gives $5\cdot 10^{-5}\,\icm$.
It would be probably necessary to approximately evaluate also $\alpha^5$\,Ry corrections 
to achieve $10^{-6}$ accuracy, as it is enhanced by the presence of $\ln\alpha^{-2}$ factors. 

At the level of accuracy of $10^{-6}\,\icm$, the proton charge radius, which contributes 
about $1.2\cdot 10^{-4}\ \icm$ to the dissociation energy of H$_2$, can be determined with 0.5\% precision, 
provided that equally accurate measurement is performed. This certainly will resolve the proton charge radius 
discrepancy, which is at the level of 4\%,  and will open a new era in precision quantum chemistry.   
Regarding the tests of hypothetical forces, which are beyond those in the Standard Model, 
the atomic scale is the natural region for the long range hadronic interactions. Moreover,
it has recently been shown that vibrational levels of the hydrogen molecule \cite{Salumbides:13}  
are particularly sensitive to the interactions beyond the Coulomb repulsion between nuclei. 
So any deviation between hopefully improved theoretical predictions and the experiment 
may signal a new physics.

\section*{Acknowledgment}

This research was supported by the National Science Center (Poland) Grants No.
2012/04/A/ST2/00105 (K.P.) and 2014/13/B/ST4/04598 (J.K.), 
as well as by a computing grant from the
Poznan Supercomputing and Networking Center, and by
PL-Grid Infrastructure.

%

\end{document}